# Boom, Bust, and Bitcoin:

# Bitcoin-Bubbles As Innovation Accelerators

Tobias A. Huber[1] and Didier Sornette[1-3]

[1] ETH Zurich, Department of Management, Technology and Economics, Switzerland

[2] Tokyo Tech World Research Hub Initiative, Institute of Innovative Research, Tokyo Institute of Technology

[3] Institute of Risk Analysis, Prediction and Management, Academy for Advanced Interdisciplinary Studies, Southern University of Science and Technology, Shenzhen, 518055, China

May 11, 2020

**Abstract**: Bitcoin represents one of the most interesting technological breakthroughs and socio-economic experiments of the last decades. In this paper, we examine the role of speculative bubbles in the process of Bitcoin's technological adoption by analyzing its social dynamics. We trace Bitcoin's genesis and dissect the nature of its techno-economic innovation. In particular, we present an analysis of the techno-economic feedback loops that drive Bitcoin's price and network effects. Based on our analysis of Bitcoin, we test and further refine the Social Bubble Hypothesis, which holds that bubbles constitute an essential component in the process of technological innovation. We argue that a hierarchy of repeating and exponentially increasing series of bubbles and hype cycles, which has occurred over the past decade since its inception, has bootstrapped Bitcoin into existence.

# 1. Introduction

The emergence of Bitcoin represents one of the most interesting technological breakthroughs and socio-economic experiments of the last decades. Not only is Bitcoin[1] a multi-faceted object, which synthesizes techno-scientific insights from various fields, it also triggered one of the largest speculative bubbles in human history. However, while the underlying cryptography, game-theory, or monetary economics, for example, have attracted scientific interest and generated an expanding academic literature,[2] a comprehensive analysis of Bitcoin as a socio-economic innovation—which situates the phenomenon in the context of the history and economics of technological innovation—has so far been lacking. In this paper, we will focus on bitcoin bubbles, in which the social dynamics driving the development and adoption of the technology crystallize themselves.

By combining various technical components, such as peer-to-peer network technology, asymmetric public-key cryptography, and a new proof-of-work algorithm, Bitcoin represents a novel, decentralized digital form of money, which substantially reduces the need for trusted third parties. As envisioned by its creator and many of his followers, it represents a radical technological innovation with potentially far-reaching socio-economic consequences. Now, irrespective of whether this vision will get realized or fail, Bitcoin offers a historically singular opportunity to advance our understanding of the nature of technological innovation and its socio-economic effects. As it is an extraordinary natural experiment—which is highly instructive but, at this scale, extremely rare in economics—the study of Bitcoin can illuminate, as we will demonstrate here, the nature of technological innovation and the social dynamics that catalyze the diffusion of emerging technologies.

Invented by a pseudonymous programmer and introduced online on an obscure mailing list, Bitcoin has since its inception experienced hyperbolic growth. Whereas the network has grown from one to an estimated 9400 nodes, the market capitalization of the protocol's cryptocurrency hyperbolically exploded from zero to almost 300 Billion USD at the peak of the so-called "crypto mania" in early 2018. As we will argue in this paper, a critical component in bootstrapping and scaling the Bitcoin network was the bubble-sequence that has driven its adoption. By accelerating the cryptocurrency's price, which, in turn, has catalyzed its speculative adoption, a series of spectacular bubbles has bootstrapped the Bitcoin network into existence. In order now to elucidate how financial speculation and excessive hype can function as necessary components in the process of developing, adopting, and diffusing novel technologies, we will dissect in the following the interwoven technological, economic, and social feedback loops that fuel Bitcoin. As we will show, Bitcoin is equally a technical as well as a social phenomenon—it represents a revolutionary technological breakthrough that will potentially have radical future socio-economic consequences. Thus, it provides a unique occasion to test and extend the Social Bubble Hypothesis, which has been developed in our research group (see Sornette, 2008; Gisler and Sornette, 2009; 2010; Gisler et al., 2011; Huber, 2017). In essence, the hypothesis, which we have refined over a series of case studies, holds that bubbles are necessary elements in the social, economic, and political processes that result in large-scale and high-impact innovations. However, while we will provide an overview of the technical properties and design of the Bitcoin network, our analysis will focus here on the socio-economic nature of Bitcoin. Consequently, the aim of the paper is threefold: i) to illuminate the technological, social, and economic dimensions of Bitcoin and their interactions; ii) to improve the hypothesis that bubbles can be phenomena with net positive benefits, as they can incubate technological and societal change, and; (iii) to extract valuable and generalizable insights from the history of Bitcoin that might help us to advance our understanding of the generic dynamics and structure of future technological revolutions.

---

[1] In this paper, we follow the convention to use uppercase "B" to refer to the Bitcoin network and lowercase "b" to the protocol-native cryptocurrency.

[2] For a sample of the academic literature, see Narayanan et al., 2016, Narayanan and Clark, 2017, Usman 2017, Wheatley et al., 2019, and Gerlach et al., 2019. However, most of the relevant analyses, on which we will rely, have occurred outside traditional academic journals on blogs or social networks, such as Medium, Reddit, Twitter, or Telegram.

While financial bubbles and market crashes have attracted abundant attention in quantitative finance and financial economics, research in these fields has mainly focused on bubbles as negative phenomena, which instantiate a form of economic inefficiency or market failure (see Allen and Gorton, 1993; Santos and Woodford, 1997; Abreu and Brunnermeier, 2003; Scheinkman and Xiong, 2003; Garber, 2000). Countering this view of bubbles as economically destructive or unproductive economic phenomena, new research has, over the last two decades, emerged outside the academic mainstream in economics that conceptualizes bubbles as important components in the process of techno-social innovation (Perez 2002; Sornette, 2008; Janeway, 2012). By analyzing Bitcoin through the conceptual prism of the Social Bubble Hypothesis, we aim to further develop our unified theory of financial bubbles and crashes (see Sornette, 2003), which provides a scientific foundation for diagnosing and predicting financial bubbles and crashes and incorporates a generic explanation for their role in technological innovation. By examining the evolution of the protocol and its cryptocurrency, we will show in more detail below that the integration of previously existing technical components and ideas into the design of the protocol—which allowed to bridge disparate and unrelated fields, methods, and concepts—enabled Bitcoin's technological breakthrough. In other words, Bitcoin provides, in our view, a valuable blueprint that can help us better understand and anticipate future technological innovations. While our analysis cannot exhaustively capture the multidisciplinary nature of Bitcoin, we nevertheless aim to identify in this paper the essential technological, social, and economic dynamics that accelerate the development and dissemination of Bitcoin.

The paper is structured as follows. In the first part, we will trace Bitcoin's historical evolution and give a synoptic view of the protocol and the technical properties of the cryptocurrency. In this section, we isolate the process of technological innovation that has resulted in Bitcoin's breakthrough. We will argue that Bitcoin instantiates a form of "combinatorial evolution," which captures the process from which novel technologies arise from the combination of preceding technological elements (see Arthur, 2009). As its historical genesis shows, Bitcoin represents an instance of "radical" or "vertical" technological innovation—as opposed to "incremental" or "horizontal" progress—precisely because its "novelty" emerged from a assemblage of existing technological components and economic incentives (see Thiel and Masters, 2012). In the next section, we will examine the social dimension of Bitcoin. By dissecting the incentive system that is embedded in the protocol, we will show that Bitcoin solves a large-scale social coordination problem by automating and formalizing social consensus between network participants and economic agents. We then analyze the cultural forces that are shaping Bitcoin's development and adoption. In particular, we identify the guiding visions driving Bitcoin's development and evangelism and the subcultures that have formed around them. Our analysis of the belief-systems that have emerged around the protocol and cryptocurrency reveals the quasi-religious dimension of Bitcoin, which manifests itself, for example, in the beliefs of some of the most committed supporters and their exegesis of Nakamoto's code and writings. As they share a structural similarity with self-fulfilling prophecies (see Merton, 1948), the beliefs that incentivize technical development and financial speculation in Bitcoin are, as we will argue, a critical factor in understanding the technology's rapid diffusion, increase in network-security, and appreciation in value. In the third section, we will further explore the self-validating nature of the Bitcoin phenomenon by mobilizing the Social Bubble Hypothesis. Based on an examination of the hype-cycles that punctuate the history of Bitcoin, we will then develop and substantiate the argument that bitcoin bubbles were necessary to bootstrap and scale the protocol and cryptocurrency. We propose that these exuberant phases need to be conceptualized as instances of *speculative technology adoption* (see also Casey, 2016). In the last section, we conclude the paper by discussing how Bitcoin can enlighten our understanding of future technological revolutions more generally.

## 2. Bitcoin: From Zero to One

On January 8, 2009, the pseudonymous programmer Satoshi Nakamoto released, on an obscure cryptography mailing list, Bitcoin—a software-protocol that allows for the decentralized transmission and storage of value. In his 2008 white paper, which provides the conceptual blueprint of the Bitcoin network, Nakamoto characterizes it as "a new electronic cash system that's fully peer-to-peer, with no trusted third party" (Nakamoto, 2008). However, the path that led to decentralized and peer-to-peer digital money is littered with failed attempts. Historically, two main currents can be identified that have coalesced into Bitcoin. Tracing back the influence on Nakamoto's protocol, these currents derive from two intertwined historical developments: cryptographic advancements in computer science and the ideologically-motivated development of cryptographically-secured, non-sovereign virtual currencies.[3] In this section, we will provide a brief outline of the intellectual history of the technical and crypto-anarchic ideas preceding Bitcoin's invention. An understanding of its pre-history will help us recognize the significance of Bitcoin as a technological as well as socio-economic breakthrough.

---

[3] This sections draws on the exceedingly thorough review of Bitcoin's academic pre-history by Narayanan and Clark (2017).

*2.1. Bitcoin: A Selective History*

Nakamoto's invention is preceded by many failed attempts to create virtual currencies, such as DigiCash, Hashcash, or Bitgold. One of the earliest and most prominent proposals is David Chaum's DigiCash. In 1989, Chaum founded DigiCash, which—by applying public key cryptography to the specific problem of digital monetary transactions— attempted to create cryptographically-secured digital cash, which emulated the properties of physical money. As early as 1983, he published the paper "Blind Signatures for Untraceable Cash" (see Chaum, 1983), which proposed so-called "blind signatures" that enabled privacy in transactions and avoided the "double-spending"-problem, which plagued many early attempts of creating digital cash. DigiCash developed a currency called ecash, which was an untraceable system of digital cash. While some banks implemented ecash, and Microsoft even proposed to integrate it into Windows, DigiCash ultimately declared bankruptcy in 1998 because of lack of merchant-adoption and support of user-to-user transactions. A wave of digital payment startups—with generic names such as CyberGold, CyberCash, or E-Gold—followed DigiCash's invention of ecash in the mid-1990s, in attempts to develop web-based money. However, except for Paypal—which pivoted away from their initial idea of enabling cryptographic payments through handheld Palm Pilot devices—these online-payment startups failed. Parallel to DigiCash and other attempts to patent and commercialize digital currencies and online-payment systems, a group of cryptographers, who interacted on what was called the "cypherpunk" mailing list, started to develop open-source alternatives (see Narayanan and Clark, 2017). While some projects, such as e-gold, proposed to peg the value of digital cash to a fiat currency or commodity, others started to experiment with free-floating digital currencies. By simulating the properties of gold, for example, some of these proposals attempted to digitally re-engineer gold's scarcity as a source of value for the native virtual currencies of these payment networks. In Bitcoin, digital scarcity has been achieved by designing a payment architecture in which the creation of money requires solving computationally expensive problems. Ideas for such systems, which Bitcoin later implemented with its proof-of-work algorithm, date back to a proposal by cryptographers Cynthia Dwork and Moni Naor, which was published in the early 1990s (see Dwork and Naor, 1992). The term *proof of work* was coined in a paper by Jakobsson and Juels in 1999 (see Jakobsson and Juels, 1999). In their paper, Dwork and Naor proposed a system in which the solution of computational problems (or "puzzles") was used to reduce email-spam. A similar idea later was expressed in Adam Back's Hashcash proposal, which he published in 1997.

In proof-of-work systems, which Hashcash and similar proposals have pioneered, the transaction validation and associated digital currency issuance—the work that needs to be proven—is performed by CPUs that invest computational recourses into a mathematical puzzle-solving exercise.[4] Although the name Hashcash already implicitly contains the idea of monetizing proof-of-work certification, the Hashcash stamps themselves, which constitute the computational proofs-of-work, were not designed to acquire monetary value. A member of the techno-libertarian cypherpunk community, Back envisioned Hashcash's proof of work-system as digital cash, and thus as an alternative to Chaum's Digicash. However, it was not possible to exchange Hashcash stamps across a peer-to-peer network. Developed after Back's Hashcash proposal, more developed proposals, which conceptualize computational puzzle solutions as digital cash, have been advanced with b-money and Bitgold (see Dai, 1998; Szabo, 2008). In both proposals, the process of solving computational puzzles is directly used for the production of digital currency. In Bitgold and b-money, which both use time-stamping to validate transactions, the computational solutions themselves instantiate monetary units. However, b-money and Bitgold, which were informally proposed on a mailing list and in a series of blog posts respectively, did not advance beyond the conceptual stage of development—they were both not implemented and lacked any code specifications.

While Nakomoto stated in 2010 on the Bitcointalk.org forum that "Bitcoin is an implementation of Wei Dai's b-money proposal on Cypherpunks in 1998 and Nick Szabo's Bitgold proposal," Bitcoin represents a technological novelty as it goes much beyond just implementing a set of pre-existing cryptographic ideas. Instead, the design of Bitcoin's architecture specifically solved deep technical and conceptual issues that earlier proposals for digital currency systems had not fully fleshed out or simply failed to address. More specifically, Hashcash, Bitgold, or b-money were all undermined by two core problems that Nakamoto's design solved: the self-monetization of the protocol-native cryptocurrency and the decentralization of network governance.

---

[4] While it is tempting to construe proof-of-work as an algorithmic reformulation of the labor theory of value—which postulates that value is determined by labor or the cost of production—"work" in Bitcoin's system is derived not from political economy but from computer science. The work to be proven—that is, transaction validations and cryptocurrency issuance performed by CPUs—is probabilistic and not deterministic in nature. In other words, no amount of computational effort guarantees a reward. Rather, the successful solution of a cryptographic puzzle is a low-probability outcome, which miners try to achieve in repeating trial-and-error-processes (see Land, forthcoming). More generally, an adequate economic framework for understanding the process of bitcoin's monetization is not the Marxist labor theory of value—elements of which can be already identified in Aristotle, Adam Smith, or David Ricardo—but the Austrian monetary economics developed by Carl Menger, Ludwig Van Mises, or F.A. Hayek. For an application of Austrian economics to Bitcoin, see Ammous (2018).

Whereas the Hashcash-system, for example, critically lacked any control of inflation, Bitcoin incorporates an automatic mechanism to periodically adjust the difficulty of the computational puzzles that regulate the issuance of new cryptocurrency. Thus, Bitcoin is capable of responding to declining hardware costs for a fixed amount of computing power, which—by substantially lowering the difficulty of producing a cryptocurrency—would result in its devaluation. By adopting an upgraded version of the Hashcash algorithm for the Bitcoin mining process, the mining difficulty adjustment—which governs Bitcoin's proof-of-work system—solved the inflation control problem, which plagued many previous digital cash proposals. In other words, as it automatically adjusts the difficulty to stabilize the rate of supply, Nakamoto was able to design a decentralized form of digital money, which removes the need for any central authority to control the inflation rate or secure the network. The monetary policy that is embedded in the protocol—which caps its supply at 21 million bitcoins—and the difficulty adjustment, which regulates the flows of energy being expended to secure the network, are, as we will show in more detail below, the source of Bitcoin's technological innovation. Furthermore, both Bitgold and b-money, for example, did not specify a consensus-mechanisms to resolve disagreement among nodes or servers about the ledger that stores all transactions in the network. Settling disagreements would have then required trusted time-stamping services for currency-creation and validation, and centralized entities controlling entry into the network to secure it from attackers attempting to alter the ledger's history or double-spend the virtual monetary units.[5] In Bitcoin, this problem is solved by the so-called "mining"-process, which was intentionally designed to be resource-intensive and computationally difficult so as to ensure that the number of mined blocks—which record transaction data—remains steady. Instead of relying on trusted servers that time-stamp transactions in a ledger—as it was proposed, for example, in a series of academic papers by Haber and Stornetta in the 1990s (see Haber and

---

[5] In a 2009 post on the P2P Foundation message board, Nakamoto states: "A lot of people automatically dismiss e-currency as a lost cause because of all the companies that failed since the 1990's. I hope it's obvious it was only the centrally controlled nature of those systems that doomed them. I think this is the first time we're trying a decentralized, non-trust-based system" (Nakamoto, 2009).

Stornetta, 1991)—Bitcoin transactions are collected by a network of untrusted "miners," which are compensated with new bitcoins and transaction fees to permanently record transaction-data into irreversible "blocks." Ordered in a linear sequence, these block give rise to what Nakamoto called "time-chain," which, later, became popularized as "blockchain" (see Nakamoto, 2009).

As this highly selective and brief history demonstrates, the design of Bitcoin synthesizes a set of existing core technical elements. Public key cryptography, Merkle Trees, cryptographic signatures and hash functions, proof-of-work, and other insights derived from the engineering of resilient peer-to-peer networks in computer science have provided the building material for the architecture of the Bitcoin network. While we are not going to delve into the intricate details of Bitcoin's technical properties[6], it is, for the purpose of this paper, sufficient to understand the network's key-components, which were incubated in its academic and cypherpunk predecessors, in order to recognize how their novel combination gave rise to Bitcoin's radical technological innovation. In order to better appreciate the breakthrough that Nakamoto's design represents, we need to zero in on the reflexive feedback loops that drive Bitcoin's security, value, and network effects. Consequently, we briefly dissect in the next section the structure of the technological as well as socio-economic incentives that are built into the protocol.

*2.2 The Techno-Economic Reflexivity of Bitcoin*

As the previous section indicates, Bitcoin represents not only a material but, in a fundamental sense, also a social technology. While the material technology that underlies the Bitcoin network consists of its codebase, the physical mining rigs, or nodes that run the Bitcoin Core software, developers, speculators, or miners, for instance, constitute the social layer of the Bitcoin architecture. Bitcoin as a social technology coordinates the behavior of this heterogeneous group of network participants that is needed for the governance of the protocol. In other words, Bitcoin's protocol governance denotes a fundamentally social process that decides upon, implements, and enforces a set of transaction and block-verification rules, which network participants can adopt. By adopting the same set of validation rules, network participants form an inter-subjective consensus about what constitutes "Bitcoin" (see Rochard, 2018). Dissenting network participants can only deviate from this inter-subjective definition of Bitcoin by "hard-forking" the protocol, that is, upgrading a copied version of the software to a new set of transaction- and block-verification rules or a different blockchain history.

---

[6] For technical treatments of Bitcoin, see Antonopoulos (2014); Song (2019).

At the core of the Bitcoin system-architecture, we can identify two reflexive components of a self-validating positive *techno-socio-economic feedback loop*, which incentivizes Bitcoin's development, valuation, and adoption (see Figure 1)[7]:

*Technological Reflexivity*: As mentioned above, specialized "miners" secure, maintain, and issue new bitcoin by competing to solve computationally intensive cryptographic puzzles. In Bitcoin's proof-of-work system, the probability of success of miners—which can be organized as companies or as mining pools—is determined by the fraction of mining power they control. If a miner successfully solves a puzzle, it gets to contribute the next "block" of transactions to the blockchain, in which blocks of transactions are linked together based on time-stamping. The mining entities are incentivized to secure and maintain the network with newly issued bitcoins. As block rewards for invalid transactions or blocks will be invalidated and rejected by the majority of miners, miners' incentives to comply with the protocol and its rules are aligned. Bitcoin's design avoids the double-spending problem, which has plagued earlier digital currency proposals, as the puzzle solutions themselves are decoupled from economic value. The amount of work required to produce a block and the amount of bitcoins issued are not fixed parameters. Rather, the block reward—which is at the time of writing (March 2020) 12.5 bitcoins/block—is programmed to halve every four years, or, approximately every 210,000 blocks (the next halving is expected to occur in May, 2020). Beyond the block reward by which new bitcoins are generated, miners are incentivized to maintain and secure the network by an additional reward scheme embedded into Bitcoin's design: senders of bitcoin payments pay miners a fee for their service of including the transaction into a block. This combined reward system that Nakamoto hard-coded into the protocol fuels, on the technological level, the reflexive feedback loop between the networks' security and value. Growth in bitcoin's value results in increased hash power allocated to the network, which, in turn, enhances its security and attracts new miners and the development and deployment of specialized mining hardware.

---

[7] As the bust of the bitcoin bubble in 2018 has shown, the positive feedback loop underlying bitcoin's rise can revert into a dynamic that results in accelerating price and devaluations accompanied respectively by increases and decreases in hash power, which, in turn, reduce network activity and security.

*Socio-Economic Reflexivity*: The reflexive incentive loop, which is built into the protocol layer and governs bitcoin's technological development, gives rise to a feedback loop on the socio-economic level. As the network's security and the native cryptocurrency's value increase, speculators, investors, and entrepreneurs are incentivized to explore and exploit the new economic space opened up by Bitcoin. Bitcoin's growth over the past decade has been driven by this self-reinforcing dynamic between discovery and speculation, which can be identified as a generic pattern in the major technological innovations of the past 250 years (see Perez, 2002; Janeway, 2013). Massive price-increases, which have been fueled by speculation, have triggered successive processes of experimentation that resulted in the gradual build-out of Bitcoin's economic infrastructure, such as the development of the second layer payment-system Lightning Network, Bitcoin-based startups that provide services such as brokerage, exchanges, wallets, public key storage, or novel alternative Bitcoin transaction systems, such as mesh or satellite networks, which are not relying on traditional Internet Service Providers.

The interlocking positive spirals of technological development and economic expectations, in turn, induce a self-validating, imitation-driven social feedback loop that accelerates Bitcoin's network effect. As past bubbles have demonstrated, exploding prices generate interest and attention, as has occurred during Bitcoin's hype cycles. Bitcoin's extraordinary returns, and the resulting media coverage and contagious virality on social media, elicit the "fear-of-missing-out" future gains inferred from extrapolations of the great gains that others have accrued in the recent past. This leads to new waves of buyers that accelerate the price growth even more. This increase in speculation and adoption, in turns, stimulates the founding and financing of new Bitcoin-based, or other cryptocurrency-related, startups that attract more speculative adopters and incentivize more Bitcoin-related research and development.

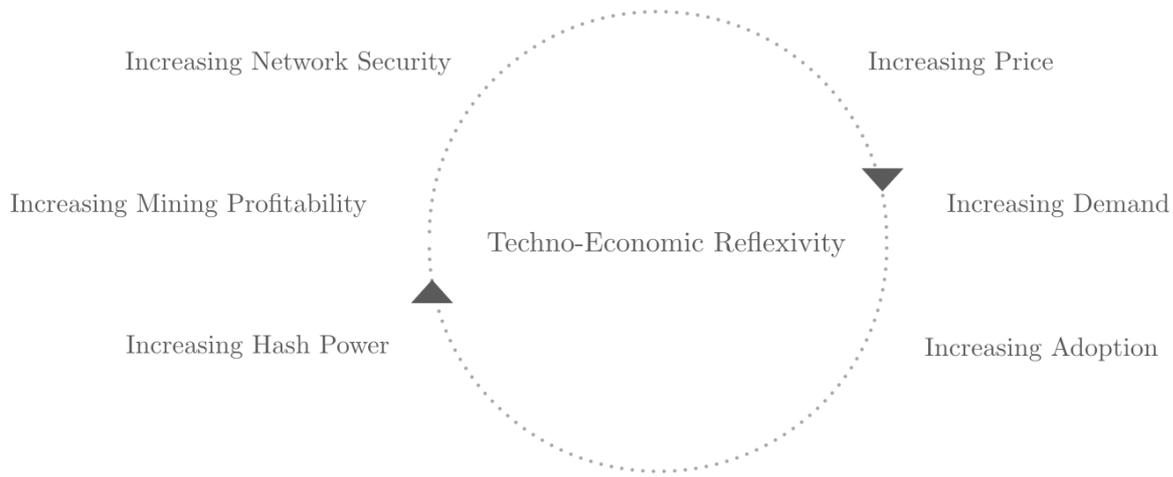

Figure 1: Bitcoin's Reflexive Techno-Economic Feedback Loop

In essence, these nested reflexive feedback loops, which constitute the incentive-system underlying Bitcoin's design, reveal the technological as well as socio-economic novelty of Nakamoto's invention. Incentivized by the scarcity encoded in the protocol, increasing demand results in more investments into specialized mining hardware and hash power allocation to the network, which, in turns, attracts more speculators, miners, and entrepreneurs that accelerate the self-validating reflexive feedback loop of security, value, and network effects.[8] In other words, Bitcoin represents a circular or closed system of socio-techno-economic incentives. As a forum post on the P2P Foundation website, dated February 18, 2009, indicates, Nakamoto himself was fully aware of the reflexive feedback loop that the design of the protocol incubates: "As the number of users grows, the value per coin increases. It has the potential for a positive feedback loop; as users increase, the value goes up, which could attract more users to take advantage of the increasing value" (Nakamoto, 2009).

*2.3 Bitcoin As A Zero-to-One-Technology*

Nakamoto's technological breakthrough does not lie in the invention of the individual components underlying Bitcoin's architecture. As we have shown above, technical and conceptual elements, such as hashing, proof-of-work, or time-stamping have existed before Nakamoto released the Bitcoin white paper (on 31 Oct. 2008), Genesis block (on 3 Jan. 2009) and code (on 9 Jan. 2009).

---

[8] In contrast to earlier Internet protocols, such as TCP/IP or SMTP, which were difficult to monetize, Bitcoin directly motivates early adopters to adopt and hype the network.

Instead, the radical novelty of Bitcoin lies in how these technical components are organized and combined in the intricate design of the protocol, which gives rise to the singular system of an automated set of technological, economic, and social incentives that have accelerated Bitcoin's rapid increase in adoption and value.[9] Bitcoin can thus be conceptualized as an assemblage of various technological components and its novelty arises from the radical re-combination of existing technologies. It instantiates an example of "combinatorial evolution," which complexity economist Brian Arthur has identified as the essential mode of technological innovation (see Arthur, 2009).[10] In other words, Nakamoto's invention represents, what venture capitalist Peter Thiel has termed, a vertical or "zero-to-one"-innovation. In contrast to horizontal innovations, which incrementally modify and improve existing technologies, zero-to one-innovations represent singularly radical new technologies (see Thiel and Masters, 2012).

Bitcoin was motivated by techno-libertarian ideals and cypherpunk beliefs in cryptographically-secured non-sovereign digital money, and therefore emerged in Internet sub-cultures (see Brunton, 2019). Given this origin, how could Bitcoin transition from a network with zero value and a single economic agent to an economic system consisting of several million users and valued at several billions dollars? How could Bitcoin become a zero-to-one technological breakthrough, with the transformative potential to disrupt the global monetary and financial system? In order to better understand how Bitcoin got bootstrapped and scaled, we need to dissect Bitcoin's underlying social dynamics and, in particular, understand the essential nature of bitcoin's bubbles. As we will show in the following sections, the bubbles that have punctuated the cryptocurrency's history do not simply instantiate a type of "collective hallucination" or "irrational exuberance" (see Odlyzko, 2010; Shiller, 2015). Rather, as we will argue, the design of the protocol itself already contains the seeds of these bubbles that have initiated the hype cycles so critical for the adoption of Bitcoin. They thus provide, as we will show, one of the purest examples of social bubbles in the sense of Gisler and Sornette (2009; 2010), Gisler et al. (2011), and Huber (2017).

---

[9] Contrary to the view that Bitcoin's underlying blockchain is the true technological innovation, we argue that the intertwined reflexive feedback loops that govern the protocol's design and incentive structure represent Bitcoin's novelty. Consequently, as it follows from our analysis, bitcoin, the cryptocurrency, cannot be separated from its underlying distributed ledger-technology as this would disrupt the intricate incentive system embedded in the network. Bitcoin commentator Joe Coin aptly captures the novelty of Bitcoin's design in a cogent blog post from 2015: "Given the crucial requirement to preserve decentralization, the problem Satoshi had to solve while designing Bitcoin was how to incentivize network participants to expend resources transmitting, validating, and storing transactions. The first step in solving that is the simple acknowledgement that it must provide them something of economic value in return [...] The incentive had to be created and exist entirely within the network itself [...] any instance of a blockchain and its underlying tokens are inextricably bound together. The token provides the fuel for the blockchain to operate, and the blockchain provides consensus on who owns which tokens. No amount of engineering can separate them" (Coin, 2015).

[10] An embryonic version of this idea can already be identified in Schumpeter's concept of "creative destruction." In 1910, he wrote that "to produce [...] means to combine materials and forces within our reach [⋯]. To produce other things, or the same things by a different method, means to combine these materials and forces differently." Schumpeter identified creative destruction as a "source of energy within the economic system which would of itself disrupt any equilibrium that might be attained" (see Schumpeter, 1934). Creative destruction precisely refers to a combinatorial process in which novelty gets continually created by combining existing elements, which, in turn, constantly disrupt the established economic order.

## 3. The Prophecy of Satoshi Nakamoto: Religion, Hype, and Technological Adoption

As can be extracted from his communications on mailing lists, Satoshi Nakamoto started to code the protocol around May 2007. After he registered the domain bitcoin.org in August 2008, he started to send out emails drafts of the Bitcoin white paper. In October 2008, he publicly released the 9 pages long white paper that specifies the protocol and, soon after, released the initial code. On January 3, 2009, Nakamoto himself mined the so-called genesis block, that is, the first 50 bitcoins. In the same year, cryptographer and early Bitcoin-developer Hal Finney, which created the first reusable proof-of-work system before Bitcoin, received the first bitcoin transaction. By December 2010, others had taken over the maintenance of the project. On December 12, 2012, Nakamoto posted his final message to the Bitcoin forum.

As it has been noted by online commentators, the genesis of Bitcoin—its beginning as an obscure and radically novel technology invented by a mysterious pseudonymous creator that has disappeared—shares structural similarities with mythologies and religion. In this section, we will attempt to systematically deconstruct the analogies between Bitcoin and religious modes of social organization. We will argue that the quasi-religiosity of Bitcoin—that is, the fact that Bitcoin adopters are often labeled as "believers," "evangelists," or "cultists"—is a basic feature of Bitcoin's technological diffusion process. In the next section, we will illuminate the structural analogy between the social dynamics governing Bitcoin adoption and religiosity. We will then outline the guiding visions that have emerged from the exegesis of Nakamoto's whiter paper and conclude by examining in more detail Bitcoin's model of technological diffusion, which is, to a large extent, based on memes and virality.

*3.1 Bitcoin As Religion*

Superficially, we can identify a few structural attributes of Bitcoin, which are analogous to religious history. As mentioned above, the most salient feature is the conceptual resemblance between Satoshi Nakamoto and religious or spiritual leaders, such as Abraham, Buddha, Jesus Christ, and their dedication towards their belief. Whereas Christ died by crucifixion as a sacrifice to achieve atonement for mankind's sins, Nakamoto sacrificed his estimated 1,148,800 bitcoins that he has never moved from the original wallet (see Lerner, 2013). Similarly, the centrality of the white paper can be analogized to a sacred scripture in organized religions. The absence of Nakamoto—often referred to as Bitcoin's "immaculate conception"—has stimulated competing exegeses of the white paper that aim to recover the true meaning of Nakamoto's messianic vision of a decentralized digital form of money. Over the past decade, incompatible interpretations of the white paper relating to technical features, such as block-size limits, have triggered a series of hard-forks. Bitcoin Cash, for example, emerged in the summer of 2017 from developers' disagreement about the block-sizes and transactions throughput. Consequently, different communities on Twitter, mailing lists, and online forums have organized around conflicting interpretations of the white paper and forks of the original Bitcoin source code, which represent the sacred object of Bitcoin. Culturally, the fragmentation into different "sects," such as so-called "Bitcoin Maximalists"—which prioritize conservative protocol development and envision it as a settlement layer for large volume payments—or "Bcashers"—which emphasize Bitcoin as a payment system—has triggered many socio-cultural conflicts. As Bitcoin full-node operators choose which vision of Bitcoin they support by running the software that enforces the protocol rules, running nodes can be reinterpreted as one of the foundational ritual practices of Bitcoin. Analogous to religions, early disciples, such as technology entrepreneur Wences Casares, who spread Nakamoto's utopian prophecy among Silicon Valley venture capitalists, are a key-ingredient in the process of diffusing technological innovation. Naturally, for some of the more radical and dogmatic believers in the original vision of Nakamoto, the creation of so-called altcoins—that is, cryptocurrencies that either directly copy Bitcoin's source code or incorporate some of its technical or conceptual properties—is, in Bitcoin's eschatology, equalized to heresy[11]. Consequently, the heresy of attempting to clone Bitcoin's "immaculate conception" requires Bitcoin Maximalists to excommunicate altcoins and their developers and supporters from Bitcoin-related forums, social media platforms, and meetups.

While the conceptual similarity between Bitcoin and religion can be dismissed as irrelevant expressions of the social dynamics that govern Bitcoin sub-cultures, it is important to emphasize the critical importance of early adopters and their excessive commitment and enthusiasm for bootstrapping novel technologies. Before examining in more detail how the process of technology diffusion is unfolding in Bitcoin, we now briefly present an overview of the most dominant visions around which Bitcoin supporters have coalesced.

*3.2 Satoshi's Vision*

Not only did the timing of the release of the white paper and software coincide with the last great financial crisis, but the message embedded in the genesis block also contained a reference to the bank bailouts occurring in 2009. As we have alluded to above, Bitcoin's history reveals a genealogical link with various techno-libertarian and cypherpunk ideals around privacy and decentralization. Indeed, Nakamoto explicitly stated in an email to Hal Finney that Bitcoin is "very attractive to the libertarian viewpoint if we can explain it properly. I'm better with code than with words though" (see Nakamoto, 2008). In other words, Nakamoto literally encoded an ideological belief-system into the base layer of the protocol, which manifests itself in the decentralized and deflationary nature of Bitcoin. As his archived communications indicate—which are littered with references to central bank policies and failures of centralized modes of organization more generally—Nakamoto envisioned the protocol as a technological alternative to centralized economic systems. Bitcoin's design thus renders explicit its inherent normative role, which, in many other technologies, remains often elusive (see Radder, 2009). It is this intrinsic ideology that has catalyzed the early proselytization of Bitcoin, which, further promoted by the succession of bubbles

---

[11] Interestingly, for theologian and philosopher René Girard, the original sin in Christianity lies in the "mimetic desire" of humans to imitate each other, which, ultimately, results in violence. On a Girardian reading, then, the emergence of altcoins, and the tribal rivalry and conflict these competing cryptocurrencies triggered, could be explained by the mimetic desire to copy the singularity of Bitcoin's design and successful implementation (see Hobart and Huber, 2019).

that unfolded from 2012 to 2017 (Gerlach et al., 2019; Wheatley et al., 2019), accelerated its early adoption. While they cannot be cleanly separated, two foundational ideological views, which drive members of the Bitcoin community, can be broadly identified:

*Bitcoin As Digital Gold*: This view, which is often inspired by Austrian Economics, emphasizes Bitcoin as a decentralized and "sound" alternative to fiat currency.[12] Given Bitcoin's finite and asymptotic money supply, supporters of this view—which, due to its monetary network effects, consider Bitcoin to be the only legitimate cryptocurrency—believe that Bitcoin represents a digital substitute for gold. In this view, its hard-coded deflationary monetary policy and decentralized design, which enables censorship-resistance and reduces the risk of confiscation, makes Bitcoin a technologically more advanced store of value that is more secure than gold and state-issued fiat-currencies. As they envision it to compete with central banks and national fiat currencies and, thus, expect massive increases in value, most Bitcoin Maximalists are committed to hoarding bitcoins—or "hodling" as it is colloquially known. Consequently, for these Bitcoin Maximalists, the protocol's primary function is not to operate as a decentralized payment-network, which competes with centralized financial services, such as the SWIFT system or PayPal. Rather, they envision the Bitcoin network as a settlement layer in which block space is used to settle large value and high volume transactions—as opposed to facilitate individual small-value individual transactions (see Ammous, 2018). Instead, it is believed that Bitcoin needs to enable small and near-instantaneous transactions on a second layer, such as the Lightning Network, which is capable of settling millions of Lightning Network payments in one finalizing transaction on the Bitcoin-blockchain. Generally, Bitcoin Maximalists, and many Bitcoin Core developers, prefer a low-rate of innovation on the base layer and favor conservative protocol-development. While they envision a gradual ossification of the base layer, innovative experimentation is however encouraged on the second layer or so-called side-chains.

*Bitcoin As Digital Cash*: As mentioned above, Bitcoin Cash is the result of a hard-fork that occurred in mid-2017. Contrary to the Bitcoin Maximalist view that Bitcoin represents digitized gold, proponents of Bitcoin Cash generally cite the subtitle and abstract of the white paper and stress that Nakamoto's initial vision was to create a borderless, peer-to-peer electronic currency. In contrast to Bitcoin, Bitcoin Cash—which increased the block-size limit so that the network can process more transactions—is designed to establish itself first as medium of exchange and not as a store of value. For ideological and technological reasons, they favor on-chain activity and are opposed to the vision that the Bitcoin network should operate as a settlement layer due to fee increases. Therefore, based on the belief that payment activity will ensure its dominance, supporters of Bitcoin Cash encourage spending instead of "hodling." Bitcoin Core implemented Segregated Witness in a soft fork, which ensures a higher degree of decentralization as it enables users to run full nodes even on low-bandwidth connections. In contrast, Bitcoin Cash supporters believe that non-mining full nodes, which only receive and validate transactions, are not relevant to the security of the protocol. However, adoption of Bitcoin Cash has failed to materialize and on-chain activity eroded.

---

[12] For a discussion on "sound money" in Austrian Economics, see Ammous (2018).

Flowing from these competing interpretations of Nakamoto's white paper are different visions of Bitcoin's future. A more moderate and pragmatic view holds that Bitcoin will adapt to regulatory constraints and integrate into the existing financial system. In this view, Bitcoin simply represents an uncorrelated asset class that can be used, similar to gold, to diversify and hedge portfolios against macroeconomic volatility and financial crises. Similarly, Bitcoin—conceived as a peer-to-peer payment network—can instantiate a decentralized and borderless alternative to centralized incumbent institutions and legacy financial networks, which are vulnerable to single-points of failures.

However, more radical futurist visions can be identified that derive from the philosophical foundations of Nakamoto's protocol, which reveal a quasi-religious set of beliefs. Believers in Hyperbitcoinization, for example, believe that the large-scale adoption of Bitcoin will result in a future demonetization of fiat currencies (see Krawisz, 2015). This belief is based on Bitcoin's censorship-resistant properties, which could undermine existing sovereign regulatory and political structures, and on superior monetary characteristics. In this view, the Bitcoin-induced collapse of fiat-currency and corresponding hyper-valuation of bitcoin, which might be triggered by systemic instabilities, such as a massive global financial crisis, would have far-reaching geo-political ramifications. Another futuristic view envisions Bitcoin as a breakthrough in information theory (see Gilder, 2018). Bitcoin, it is assumed, could serve as platform for general-purpose computation. Others even compare the Bitcoin network to a collective self-organizing intelligence (see Greenhall, 2016) or a "new form of life."[13]

Now, disregarding the plausibility and probability of such futuristic scenarios of Bitcoin adoption (see Senner and Sornette, 2019, for a critical review), these beliefs convey the Messianic dimension of Nakamoto's writings, which, for many proponents, promise technological salvation. However, it is precisely this set of philosophical beliefs, based on the different interpretations that we have outlined above, which has culturally accelerated the adoption of Bitcoin. In the next section, we will thus briefly describe Bitcoin's technological adoption cycles before we analyze in more detail how speculative Bitcoin bubbles accelerate the process of technology diffusion.

*3.3 Bitcoin Evangelism and Technology Diffusion*

Bitcoin has followed a specific technology adoption cycle that has been modulated by different social forces. Since its invention, we can discern four distinct, albeit idealized, phases in Bitcoin's diffusion. These phases of the ongoing technological adoption cycle have corresponded to bitcoin's price-acceleration bursts. In other words, each burst in bitcoin's price attracted a new set of adopters and resulted in a more widespread diffusion of the technology.

Bitcoin's technology adoption cycle has been initiated by a small group of believers, which constitutes the first cohort of adopters (see also Boyapati, 2018). Their adoption of this bleeding edge technology, which was motivated by the techno-ideological reasons illuminated earlier, elicited a process of continuous experimentation, debugging, and testing that gradually stabilized and improved the Bitcoin Core software. The quasi-religious devotion and extreme enthusiasm of this cohort of adopters, which consisted mainly of cryptographers, cypherpunks, and developers, then infected a group of ideologically motivated technologists, investors, and technology entrepreneurs, who in turn started to evangelize Bitcoin. The gradual build-out of the Bitcoin infrastructure and the first rudimentary exchanges, such as the infamous Japan-based Mt. Gox exchange, which allowed the conversion from fiat currency into Bitcoin, attracted early retail investors, which define the third cohort of adopters. The liquidity that these early speculators provided resulted in the first large-scale Bitcoin bubble in 2013, which triggered an inflow of more capital and attention. The launch of regulated exchanges, such as GDAX or Bitstamp, and OTC brokers, such as Cumberland Mining, in turn initiated the ongoing institutionalization of Bitcoin and intensified its virality.[14] If we now map Bitcoin's diffusion onto the generic technology adoption cycle, the fourth phase started after the bear market that followed the burst of the 2013 bubble, which lasted from 2014 to 2016. This phase of adoption is marked by the entry of the "early majority" of retail and institutional investors. Accelerated by the formation of regulated futures markets, such as the CME and CBOE, and other exchange-traded products, the price of bitcoin increased to almost 20'000 USD in December 2017. The infrastructure, which has been build out during the last bubble, might in the future usher the "late majority" and "laggards" phase of the technology adoption cycle.[15] Given that speculative frenzies boosted Bitcoin's adoption, in the next section, we provide a more granular analysis of how bubbles have catalyzed the adoption of the cryptocurrency.

---

[13] For example, cryptographer Ralph Merkle, who invented Merkle Trees—a data structure that Bitcoin employs—compares the protocol to an organism: " [...]. Bitcoin is the first example of a new form of life. It lives and breathes on the internet. It lives because it can pay people to keep it alive. It lives because it performs a useful service that people will pay it to perform. It lives because anyone, anywhere, can run a copy of its code. It lives because all the running copies are constantly talking to each other. It lives because if any one copy is corrupted it is discarded, quickly and without any fuss or muss. It lives because it is radically transparent: anyone can see its code and see exactly what it does. It can't be changed. It can't be argued with. It can't be tampered with. It can't be corrupted. It can't be stopped. It can't even be interrupted [...]. But as long as there are people who want to use it, it's very hard to kill, or corrupt, or stop, or interrupt" (see Merkle, 2016).

---

[14] Bitcoin's diffusion occurred primarily online on social media, mailing lists, and blog posts. For example, Bitcoin's infectiousness has spread with various "memes," which acts as a unit for carrying and transmitting Bitcoin-related ideas and symbols. An example of Bitcoin's mimetic model of technology diffusion is the "Hodl"-meme, which—resulting from a misspelling of the word "hold"—motivates bitcoin holders to resist the urge to sell in response to market fluctuations. Consequently, analysts at Barclays developed an epidemiological model of Bitcoin's diffusion that models bitcoin as a "virus" that "infects" the population adopting the cryptocurrency technology.

[15] In Perez's classic conceptual model of technology diffusion, this phase might correspond to what she identifies as the "turning point." In her model, each technological disruption is triggered by a financial bubble, which allocates excessive capital to emerging technologies. Perez has extracted a regular generic pattern of technology-diffusion from historical case studies. She identifies an "installation"-phase in which a bubble drives the installation of the new technology. This is followed by the collapse of the bubble or a crash, to which she refers to as the "turning point." After this transitional phase—which occurred, for example, after the first British railway mania in the 1840s, or, more recently, after the dotcom-bubble—a second phase is unleashed: the "deployment" phase, which diffuses the new technology across economies, industries and societies (see Perez, 2003).

## 4. Bitcoin-Bubbles As Innovation-Accelerators

*4.1 A Brief Overview of Bitcoin Bubbles*

Bitcoin's history is punctuated with speculative bubbles (Gerlach et al., 2019). Since the inception of the first exchange-traded price in 2010, the technological diffusion of Bitcoin can be conceptualized as a series of boom-bust cycles of increasing intensity and magnitude. This sequence of super-exponentially accelerating price-increases that are followed by equally spectacular crashes seems to follow the path of the classic Gartner Hype Cycle, which is used as a generic representation of the different phases of technology adoption. These hype cycles, which we will analyze in more detail below, have been fueled by speculate bubbles that, in turn, have generated more widespread diffusion of the technology. Consequently, each cycle corresponds to the distinct phases of adoption highlighted in the previous section.

We can identify five bitcoin-bubbles (see Figure 2) (see also Wheatley et al., 2019). In 2011, bitcoin's price increased from 1 USD on April 14 to 28.90 USD on June 9. In the following year, the price increased from 4.80 USD, on May 10, to 13.20 USD on August 15. In 2013, from January 3 to April 09, the price of bitcoin increased from 13.40 USD to 230 USD. In the same year, bitcoin increased from USD 123.20 on October 7 to USD 1156.10 on December 4. After the price crashed at the end of 2013, the price slowly recovered over a period of two years. On March 25, 2017, bitcoin's price started to accelerate from 975.70 USD to 20.089 USD on December 17, 2017, which represents bitcoin's all-time high. As this pattern of recurring bitcoin-bubbles demonstrates, each crash or correction was followed by an even larger-bubble in absolute prices (but of similar and very large amplitudes when measured in relative price changes). Bitcoin's price during the aforementioned bubbles was largely correlated with an increase in liquidity and with the maturation of the infrastructure, which attracted new adopters, such as entrepreneurs or speculators. While it was exceedingly difficult to trade bitcoin during the first bubble—which were primarily acquired through mining—exchanging and securing bitcoin has become relatively easy during the bitcoin-bubble that peaked in December 2017.

Bitcoin's price can thus be characterized by a hierarchy of repeating and exponentially increasing bubbles (Gerlach et al., 2019). These bubbles represent phases of unsustainable accelerating phases of price corrections and rebounds, which are driven by self-reinforcing feedback loops of herding behavior (Sornette, 2017). While the collapse of prices, which follows the faster-than-exponential power law growth processes defining bubble regimes (Sornette and Cauwels, 2015), can be destabilizing and destructive, bubbles of this type need to be understood as a source of technological innovation. By attracting capital in excess to what would be justified by a rational cost-benefit analysis or by a standard discounted cash flow calculation, bubbles accelerate the development of emerging technologies and, as Bitcoin clearly demonstrates, technology adoption cycles. Capital flows in at the early stage, which leads to a first wave of price increases. Attracted by the prospect of extrapolated higher returns, more investors follow, which triggers a positive feedback mechanism that fuels spiraling growth. Bubbles, which have historically incubated major technological innovations and disruptions, share a central dynamic: the funding of these new technologies decouples from rational expectations of economic return and, correspondingly, result in a reduction of collective risk-aversion. Irrespective of quantifiable financial returns and economic values, bubbles mobilize the financial capital needed to develop new transformative technologies. Based on the observation that these bubble dynamics extend beyond financial markets to social systems, one of the authors has, in a series of detailed case studies, developed the Social Bubble Hypothesis (see Sornette, 2008; Gisler and Sornette, 2009).

Based on the insight that many large-scale technological, social, or political projects involve collective enthusiasm and over-optimism, which results in unrestrained investment and commitments as well as a general reduction of risk aversion, the Social Bubble Hypothesis provides a useful conceptual framework to understand the emergence of Bitcoin.[16] In the next section, we will analyze the series of bitcoin bubbles identified above through the lens of the Social Bubble Hypothesis.

---

[16] Mencius Moldbug, a pseudonym for technology entrepreneur and computer scientist Curtis Yarvin, embodies the spirit of the bubble dynamics of bitcoin, with his Bubble Theory of Money (BTM), which holds that, given its fundamentally social nature, money can be likened to a bubble. He writes: "Bitcoin is money and Bitcoin is a bubble, The BTM asserts that money and a bubble are the same thing. Both are anomalously overvalued assets. Both obtain their anomalous value from the fact that many people have bought the asset, without any intention to use it, but only to exchange it for some other asset at a later date. The two can be distinguished only in hindsight. If it popped, it was a bubble. If not, money—so far"

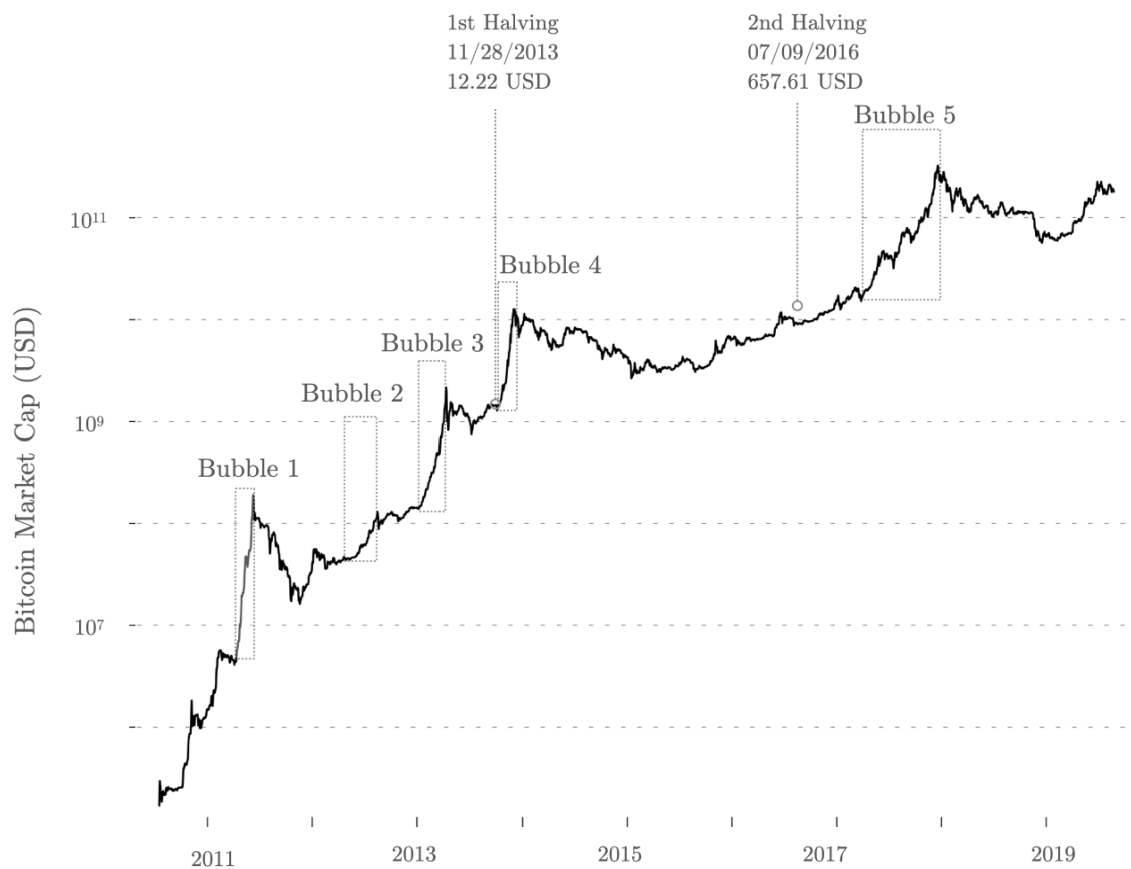

Figure 2: Main Bitcoin Bubbles and Hype Cycles. Notice the vertical scale and the larger than tenfold price increase in less than a year in each of these bubble episodes. Gerlach et al. (2019) document these bubbles as well as many other smaller ones covering this period.

*4.2 Bitcoin As A Social Bubble*

Similarly to financial bubbles, the essential ingredients of social bubbles are socio-behavioral mechanisms, such as herding or imitation, exuberant over-optimism and unrealistic expectations. By generating positive feedback cycles of extraordinary enthusiasm and investments—which have

---

(see Moldbug, 2013; Law, 2006). This reasoning should be distinguished from the standard theory of money, which considers it as an IOU and thus as credit (von Becke and Sornette, 2017), and the fact that credit growth is unstable and has led historically to boom-bust cycles over the last 5000 years (Graeber, 2012). In the creation of bitcoins, there is indeed no credit mechanism. The BTM is also reminiscent of the theory of value considered as a convention, developed by the French economist André Orléan (1987; 1989).

been essential for bootstrapping various social and technological enterprises—speculative bubbles can accelerate technological innovation. The complex networks of social interactions between enthusiastic supporters have catalyzed the formation of many large-scale scientific or technological projects. Characterized as manifestations of collective over-enthusiasm, they constitute an important element in the dynamics that give rise to scientific discoveries and radical technological breakthroughs. Similar to the generic technology hype cycle discussed previously, social bubbles are initiated by a burst of enthusiasm for a new technology. The earliest adopters and investors have strong convictions about the transformative nature of the technology they are investing in. The unbridled enthusiasm and commitments result in accelerated prices, which, in turn, catalyze more investments and speculation. Eventually, enthusiasm and investments peak, and the cycle is exhausted and prices and commitment saturate or decrease.

In the case of Bitcoin, in the early phase of the social bubble, the over-enthusiasm, commitment, and strong social interactions of cryptographers, computer scientists and cypherpunks significantly fueled the development and adoption of the technology. The enthusiasm and commitment of the cohort of early Bitcoin adopters then triggered the interest of early speculator and investors, which were often ideologically motivated to invest in the technology. It was this flow of capital and interest that triggered the first bitcoin bubbles in 2012 and 2013. After the peak of the first large bitcoin bubble, when bitcoin reached for the first time a price of more than 1000 USD in November 2013, the bubble collapsed and interest decreased substantially. The speculative fervor, which gave rise to super-exponential price growth, was then followed by despair, public derision, and a sense that the technology was not transformative at all. Eventually, bitcoin's price bottomed and went through a plateau during which a cohort of new believers and investors became attracted by the importance of the technology. Bitcoin's price-plateau persisted for two years before a new bubble gradually started to form in 2015. Over the prolonged bear market that lasted from 2013 to 2015, a new base of adopters has formed for the next iteration of the bubble cycle. This next iteration of the bubble, which in 2017 resulted in unprecedented hype and attention, attracted a much larger number of adopters. The cycle of bitcoin bubbles, which has given rise to accelerating prices and increasing media attention, has woven a network of reinforcing feedback loops that have led to widespread over-enthusiasm and commitment among Bitcoin Core developers, entrepreneurs, or speculators. This has been fueled by excessive expectations of ever-increasing price-acceleration and technology adoption. This momentous enthusiasm, which for instance led early cypherpunks and technologists to test and improve Bitcoin's code, and the extremely high expectations and hype towards the transformative potential of Bitcoin, constitute essential elements in the dynamics of Bitcoin's development and diffusion. In the next section, we will examine the nature of Bitcoin's sequence of bubble-driven hype cycles in more detail.

*4.3 A Hierarchy of Bitcoin Hype Cycles, Speculative Bubbles, and Technological Adoption*

Bitcoin's technological adoption—which could also reflect the monetization process of Bitcoin (see Boyapati, 2018)—seems to follow a hierarchical pattern of speculative bubbles within speculative bubbles that matches the shape of the classic Gartner hype cycle. The hype cycle, which represents the adoption of emerging transformative technologies, distinguishes between five phases. In the first phase, a technological breakthrough triggers initial interest. This phase corresponds to Nakamoto's release of the Bitcoin white paper and software, which attracted technologists and cypherpunks, such as Hal Finney who started to experiment with the Bitcoin technology when it was still in its proof-of-concept stage. Early adopters then started to improve the Bitcoin software. The first spike in Bitcoin's price occurred on July 12, 2010 on the first bitcoin exchange, The Bitcoin Market, after an article about Bitcoin Version 0.3 appeared the day before on the popular technology website site *Slashdot*. Following the launch of the Mt. Gox exchange in July 2010, bitcoin price peaked in June 2011 at 31.90 USD. During this first phase, Bitcoin entered the "Peak of Inflated Expectations" on the hype cycle. However, as the price of bitcoin decreased by over 93% over the following four months, Bitcoin entered the "Trough of Disillusionment," which is characterized by decreasing interest. After the price bottomed in April 2013, another price spike passed the psychological resistance level of 100 USD, which was fueled by the financial crisis in Cyprus that boosted bitcoin demand due to the growing distrust of banking from the threat of confiscation of banking deposits. However, after reaching a new high of 266 USD on Mt. Gox, it soon crashed below 60 USD before slowly returning to the range of 120 USD. After a longer phase of price stabilization, speculative investment resumed and Bitcoin entered the "Slope of Enlightenment," in which entrepreneurs have launched new Bitcoin-related startups and products. The next phase of technology adoption, which Bitcoin has not entered yet, is the "Plateau of Productivity" that is characterized by large-scale mainstream adoption.

While the trajectory of bitcoin's price since its inception can be mapped onto a generic hype cycle, it is important to note that each speculative bubble itself follows the path of a hype cycle. In other words, Bitcoin's technological adoption can be conceptualized as series of nested hype cycles, with a hierarchy of magnitudes and time scales. Unlike the generic Gartner hype cycle, however, Bitcoin's volatile curve of adoption does not follow a steady gradual increase. Instead, as Bitcoin's adoption has been speculative in nature, it followed a sequence of even more extreme growth phases than the standard exponentially growth path of the S-curve, which ended in a series of spectacular crashes. Generically, the hierarchical pattern of Bitcoin hype cycles seem to result from herding and imitation behavior of traders, which gives rise to speculative bubbles. As price accelerates, more speculators start to buy bitcoin. Eventually, as prices increase even more, early speculators are driven to take profits, which then triggers a correction or crash. Consequently, after each bubble-crash sequence, in which new long-term investors, or so-called "hodlers," are attracted, the amount of long-term holders increases during the bubble component of the cycle. In other words, due to its speculative bubbles, Bitcoin has been able to continually expand its adopter base.

Bitcoin's bubble-fueled technological adoption cycles can thus be conceptualized as a pattern of nested curves that each represent a new cohort of "hodlers." These subsequent waves of new "hodlers"—which represent future speculators who are not willing to sell in the next crash—can be quantified with a Bitcoin-native accounting structure called an UTXO—an "Unspent Transaction Output" (see Bansal, 2018). UTXO's, which are time-stamped by the transaction/block in which they were created, represent when a bitcoin was last used in a transaction. We can identify different adoption waves since Bitcoin's release, which occur when a cohort of new speculators or investors buy bitcoins during a bubble and hold through the downturn into the next market cycle. Visually, in Figure 4, these speculative adoption waves can be represented by different age bands: whereas warmer-colored age bands (<1 day, 1 day–1 week, 1 week–1 month) represent transactions of large amounts of bitcoin, the steady growth of the top, cooler-colored age bands (2–3 years, 3–5 years, >5 years) indicate the adoption of Bitcoin, that is, they represent an increase in "hodling."[17] These adoption waves manifest themselves visually as nested curves, which are caused by each age band becoming progressively wider (see Figure 4). The different levels of unspent transaction outputs indicate that each bubble attracts a new cohort of "hodlers" who are accumulating and holding bitcoin. In other words, each speculative bubble has triggered a "hodling" wave. Bitcoin thus represents one of the purest examples of *speculative technological adoption*. The bubble-driven repeating and super-exponentially increasing hype cycles continually attracted new cohorts of "hodlers.

---

[17] "Hodling" represents the speculative adoption of bitcoin as it implies a speculative bet on future gains in value. Thus, adoption is here defined as an increase in "hodling"—that is, the accumulation and holding of bitcoin.

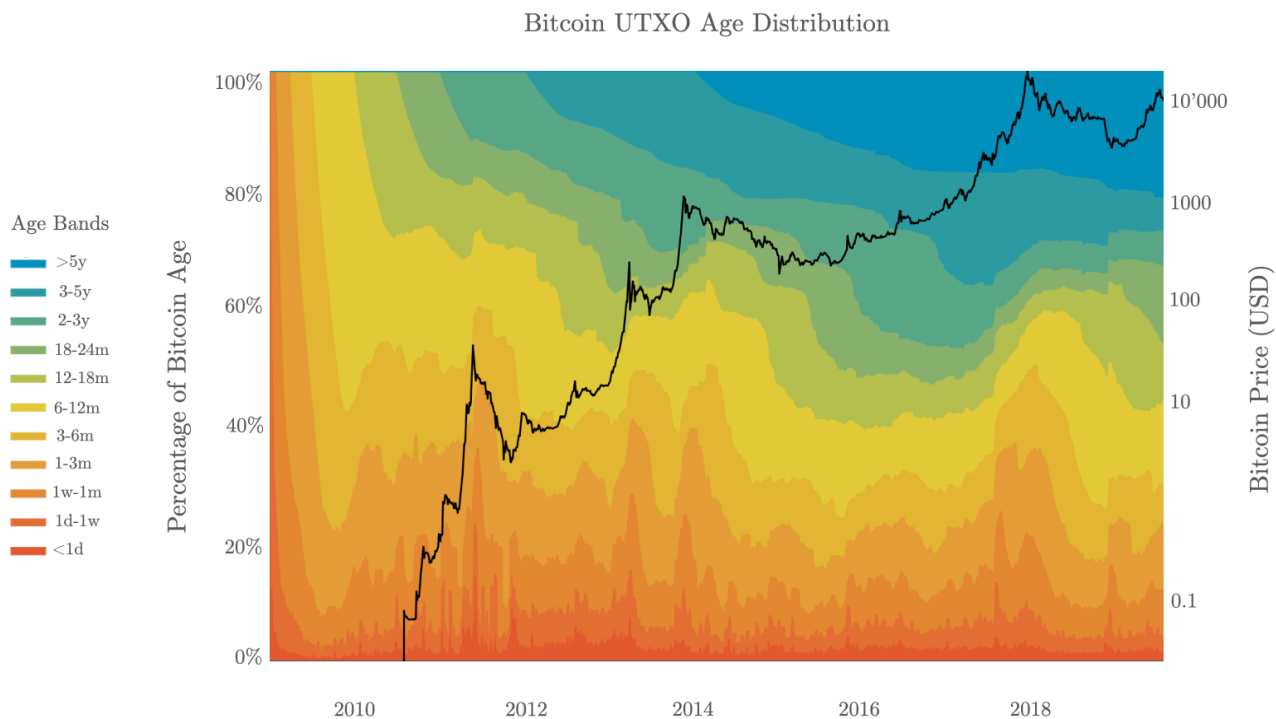

Figure 3: Speculative Bitcoin Adoption Waves in colors (left axis and color codes) superimposed on the Bitcoin price (black line and right axis) (adopted from Bansal, 2018).

*4.4 Bitcoin's Technological Adoption Cycle*

It is interesting to note that the hype-cycles fueling Bitcoin's technological adoption are embedded in the protocol itself. In Nakamoto's design, as mentioned above, every four years a halving occurs that reduces the reward for miners by half. Built into the protocol to control Bitcoin's inflation, the previous so-called halvening have coincided with massive price-accelerations. After the first halvening, which occurred in November 2012, Bitcoin's price increased from 12 USD to more than 650 USD at the time when the second halvening in July 2016 occurred. After the block reward reduction to 12.5 bitcoins, where each block is created every 10 minutes, the price accelerated to almost 20'000 USD. While it is of course uncertain whether the next halvening —which will occur in mid-2020—will accelerate prices in a similar way, the previous halvening have fueled the main

Bitcoin's hype cycles. As his message mentioned above indicates, Nakamoto seemed to have programmed speculative bubbles into the protocol, with the intention of accelerating the feedback loops needed to bootstrap bitcoin's value. In the message posted on the P2P Forum in 2009 Nakamoto stated: "As the number of users grows, the value per coin increases. It has the potential for a positive feedback loop; as users increase, the value goes up, which could attract more users to take advantage of the increasing value" (Nakomato, 2009). Positive feedback is known to be the main mechanism for the generation of bubbles (Sornette, 2017; Johansen and Sornette, 2010; Jiang et al., 2010; Sornette and Cauwels, 2015). It is thus plausible that Nakamoto designed the halvenings to create "artificial" boom-and-bust cycles. The 4-year-halvening cycles drive up prices, which are then followed by an increase in hash rate and number of hodlers. Even after a crash, the hash rate and, consequently, the security of the network are higher than before the price-acceleration. Moreover, the halvening also attracts new cohorts of adopters to "hodl."[18] Over the past decade, this sequence of speculative bubbles thus bootstrapped a new form of digital money, which started with zero value (and arguably with zero fundamental value in the standard economic sense) to a network that, at the peak of the last bubble, was valued at more than USD 320 billion. Rather than simply representing excessive speculation, we can conclude that bubbles and hype cycles have been accelerating Bitcoin's technological adoption. In other words, speculative bubbles provide the fundamental mechanism fueling the intertwined techno-economic feedback loops that drive Bitcoin's security, value, and network effects.

## 5. Conclusion and Discussion

We have elaborated on the elements supporting the claim that Bitcoin represents a technological breakthrough. As we have noted, Bitcoin represents a radical technological innovation, not simply because it represents a novel technology. Rather, the novelty emerges from Nakamoto's combination of ideas and technologies that existed previously in disparate and previously unrelated fields. As its historical genesis shows, Bitcoin represents an instance of "radical" or "vertical" technological innovation—its "novelty" emerged from an assemblage of existing technological components. The invention of Bitcoin thus required the bridging of disparate fields, terminologies, and assumptions. Bitcoin's breakthrough lies in how Nakamoto designed a system of interlocking techno-economic feedback loops that fuel its value, security, and network effects.

---

[18] The expectation that the halvening results in an increase of "hodlers" and higher prices assumes that scarcity drives bitcoin's value. In this view, the halvening represents a supply "shock" that increases bitcoin's relative scarcity. While the supply cap of 21 million bitcoins is algorithmically fixed, relative supply and the bitcoins in circulation decrease. Furthermore, due to the change in the supply schedule, the market needs to absorb fewer bitcoins, which miners are selling to cover their capital expenditures. Gradually, miner compensation will transition to transaction fees. A popular model that is used to model bitcoin's scarcity-based value is the so-called Stock-to-Flow model. It models the price of Bitcoin based on the "stock-to-flow ratio," which was initially used to value gold and other raw materials. By relating the "stock"—i.e., the quantity issued–to the "flow"—i.e., the annual issued quantity–the model derives a prediction of a bitcoin price post-halvening of $55,000 to $100'000 (which would correspond to a market cap of more than $1 trillion) (see Plan B, 2019).

Bitcoin genesis is different from any technological breakthrough that has historically preceded it. As we have documented above, Bitcoin's "immaculate conception" by a pseudonymous programmer has given rise to a community of developers and users who have a quasi-religious commitment to the cryptocurrency. Given Bitcoin's open-sourced design and distributed architecture, the technology represents a permissionless and decentralized model of innovation. Bitcoin did not result—as it was historically the case with preceding technological breakthroughs—from a specific set of innovation policies or government-funded academic research. Rather, it emerged outside the boundaries of academic peer-review or government-funding. It appeared on an obscure mailing list, which was adopted and diffused by a group of extremely committed and enthusiastic supporters. In this sense, Bitcoin and the self-organizing principles that govern the protocol's evolution share an essential similarity with the emergent properties that complex systems exhibit, which Hayek characterized as "spontaneous order" (see Hayek, 1969). It will be interesting to observe whether the beliefs of these enthusiastic supporters, which evangelize Bitcoin, will in the future follow the trajectory of a self-fulfilling prophecy that will continually attract new developers, entrepreneurs, and "hodlers" as it has been the case until now. Furthermore, as we have shown above, what is singularly unique in Bitcoin is that hype-cycles are built into the design of the protocol itself. The deflationary nature of the cryptocurrency's supply and halving of block-rewards have triggered a process of what can be called *speculative technology adoption*. The resulting hierarchical sequence of repeating and super-exponentially increasing series of bubbles, which have occurred over the past decade since Bitcoin's inception, has resulted in new waves of speculative adopters.

As we have argued, these speculative bubbles and hype-cycles have bootstrapped the Bitcoin network into existence. Whereas financial bubbles have historically been important catalysts in the diffusion of technological revolutions, Bitcoin represents the first radical technological innovation in which bubbles constitute necessary components in the process of technology adoption and diffusion. As our previous discussion of the halvening-cycle has shown, the emergence of bubbles and hype-cycles, which accelerate the technological adoption, are hard-coded into the protocol. Yet, the fundamental question remains of whether Bitcoin's technological breakthrough can be replicated. Can Bitcoin's "immaculate conception"—that is, its invention by a pseudonymous programmer, which attracted a following of committed believers—get repeated? While Bitcoin's invention represents a technological singularity,[19] its history nevertheless demonstrates the importance of hype and bubbles for the development and diffusion of cutting-edge technologies. A more generalizable insight, which can be derived from our Bitcoin case study, is that bubbles need to be conceptualized as "chaotic attractors" for technological innovation. The development of a definite vision of the future—which generates hype and great enthusiasm that reduces risk-aversion and attracts new supporters and adopters—is driving the Bitcoin phenomenon. While the Bitcoin experiment is still unfolding, its emergence has clearly demonstrated that hype and bubbles constitute essential elements in the process of technological innovation. In order to be successful, future technological innovations, we suggest, will thus need to incubate and generate comparable visions, enthusiasm, and hype, which are currently fueling the dynamics of Bitcoin's development and diffusion.

---

[19] Bitcoin could only be invented once. Given its singular nature, it has been argued that Bitcoin cannot be replaced by another cryptocurrency. Hal Finney, for example, stated that every subsequent version of the protocol designed to substitute Bitcoin would be self-invalidating. A hypothetical Bitcoin successor would undermine its own viability and credibility as "an investor" would not "know that it won't happen again" (Finney, 2011). In this view, the adoption of Bitcoin follows a binary logic: either Bitcoin succeeds or Bitcoin and all other forks or cryptocurrencies will fail as well.

Combining a Generalised Metcalfe's Law and the LPPLS Model, *Royal Society Open Science* 6, 180538, (http://dx.doi.org/10.1098/rsos.18053), 1-13 (2019).